\documentclass[11pt]{article}

\usepackage{geometry}  
\usepackage[T1]{fontenc}  
\usepackage[utf8x]{inputenc}  
\usepackage{fancyhdr}  
\usepackage{enumitem}  
\usepackage{hyperref}  
\usepackage{titling}  
\usepackage{natbib}  
\usepackage{mathtools}  
\usepackage{titlesec}  
\usepackage{lastpage}  

\usepackage[english]{babel}  
\usepackage{amsmath}  
\usepackage{amsfonts}  
\usepackage{amssymb}  
\usepackage{wasysym}  
\usepackage{bbm}  
\usepackage{array}  
\usepackage{xr}  
\usepackage{verbatim} 
\usepackage{float} 
\usepackage[ruled,vlined]{algorithm2e}
\usepackage{subcaption}
\usepackage{booktabs}
\usepackage{pgfplots}
\pgfplotsset{compat=1.18}
\usepackage{subcaption}
\usepackage{multirow}

\hypersetup{%
  pdfborderstyle={/S/U/W 1}
}

\setcitestyle{round,numbers}
\titleformat{name=\section}{\normalfont\Large\bfseries}{}{0pt}{}
\titleformat{name=\subsection}{\normalfont\large\bfseries}{}{0pt}{}
\titleformat{name=\subsubsection}{\normalfont\normalsize\bfseries}{}{0pt}{}

\newcommand{\email}[1]{\href{mailto:{#1}}{{#1}}}

\newcommand{\keywords}[1]{\textbf{Keywords}: {#1}}

\newcommand{\optincludegraphics}[2][]{}

\newcommand{\optinput}[1]{}

\newtagform{brackets}{[}{]}
\usetagform{brackets}

\newcommand{\thejournal}[1]{Magnetic Resonance in Medicine}

\title{Quantum Adaptive Sensing for Accelerated MRI}

\immediate\write18{texcount -sum=1,1,1,1,1,1,1 -merge -incbib -dir -utf8 \jobname.tex > \jobname.wcTotal }
\immediate\write18{texcount -sub=none -merge -incbib -dir -utf8 \jobname.tex | grep _main_ | grep -oE '[0-9]+' | python -c "import sys; print(sum(int(l) * w for l, w in zip(sys.stdin, [1, 1, 0, 0, 0, 0, 0])))" > \jobname.wcManuscript }
\immediate\write18{texcount -sub=none -merge -incbib -dir -utf8 \jobname.tex | grep Abstract | grep -oE '[0-9]+' | python -c "import sys; print(sum(int(l) * w for l, w in zip(sys.stdin, [1, 1, 0, 0, 0, 0, 0])))" > \jobname.wcAbstract }

\newcommand{\wcTotal}{\clearpage{\noindent\large{\bf Detailed Word Count} (not to be included for submission)}\verbatiminput{\jobname.wcTotal}}
\newcommand{\wcManuscript}{\input{\jobname.wcManuscript}}
\newcommand{\wcAbstract}{\input{\jobname.wcAbstract}}

\begin{document}

\begin{titlepage}
{\noindent\LARGE\bf \thetitle}

\bigskip

\begin{flushleft}\large
	Asmit Ganguly\textsuperscript{1},
	Suprajit Dewanji,
	Chenyang Zhao\textsuperscript{1},
	Danny J. J. Wang\textsuperscript{1,{*}}
\end{flushleft}

\bigskip

\noindent
\begin{enumerate}[label=\textbf{\arabic*}]
\item Laboratory of Functional MRI Technology, Neuroimaging and Informatics Institute, University of Southern California, Los Angeles, United States

\end{enumerate}

\bigskip


\textbf{*} Corresponding author:

\indent\indent
\begin{tabular}{>{\bfseries}rl}
Name		& Danny J. J. Wang												\\
Department	& Biomedical Engineering													\\
Institute	& University of Southern California														\\
Address 	& 2025 Zonal Avenue														\\
			& 90033 Los Angeles														\\
            & United States														\\
E-mail		& \email{dannyjwa@usc.edu}											\\
\end{tabular}

\vfill



\end{titlepage}

\pagebreak

\begin{abstract}

Compressed sensing accelerates magnetic resonance imaging (MRI) by reconstructing images from undersampled k-space measurements, but reconstruction fidelity depends strongly on the spatial distribution of the acquired samples. We introduce an adaptive k-space sampling framework that formulates sequential Cartesian phase-encode-line selection as a fixed-cardinality quadratic unconstrained binary optimization (QUBO) problem. The objective dynamically combines prior preference for central k-space, signal-energy information obtained from previously acquired measurements, and pairwise interactions that promote spatially dispersed sampling. This formulation is compatible with both classical annealing algorithms and quantum annealing hardware, enabling current evaluation with scalable classical solvers while providing a pathway toward accelerated optimization on future quantum processors with greater problem capacity and connectivity. The principal experiments were conducted retrospectively using simulated eight-coil three-dimensional MRI data, with the QUBO optimized using parallel tempering and the resulting measurements reconstructed using SENSE with total-variation regularization. At both 20\% and 10\% sampling, the proposed method improved PSNR, SSIM, NMSE, and HFEN relative to the evaluated static Cartesian sampling strategies, including variable-density Poisson-disc sampling, with the magnitude of improvement depending on resolution, acceleration, and noise level. In a separate reduced-pool experiment, implementation with a D-Wave quantum-classical hybrid solver produced reconstruction quality comparable to variable-density Poisson-disc sampling, demonstrating the feasibility of deploying the formulation on current quantum optimization infrastructure. Although the present experiments do not establish quantum computational advantage, the direct QUBO representation positions the method to exploit advances in quantum-annealing hardware that may reduce decomposition and optimization overhead for larger adaptive sampling problems. These findings establish adaptive QUBO optimization as a promising framework for accelerated MRI and motivate prospective scanner validation and systematic quantum–classical benchmarking.

\end{abstract}

\bigskip

\pagebreak

\section{Introduction}

Magnetic resonance imaging (MRI) stands out as one of the most powerful techniques for non-invasive imaging of the human body and its activities in real time. The signal acquisition process involves spatial encoding of nuclear spins using linear magnetic field gradients ($G_x, G_y, G_z$), which map the spatial distribution of protons onto the Fourier spatial-frequency domain, known as k-space, representing the 2D or 3D Fourier transform of the magnetization being imaged.

In conventional Cartesian acquisition schemes, k-space is sampled through a series of phase-encoding steps. The total encoding time depends on the phase-encoding steps ($N_y$) and the repetition time ($T_R$). To achieve high spatial resolution without aliasing, a large number of samples must be collected, satisfying the Nyquist-Shannon criterion. This requirement often leads to prohibitively long scan durations, which increases susceptibility to motion artifacts, limits patient throughput, and makes it difficult to capture rapidly changing physiological processes. To reduce acquisition time, compressed sensing \cite{cs_optimization} of under-sampled k-space has become a popular method for MRI acquisition. Multiple undersampling techniques have been studied using trajectories such as Radial, Spiral, random, and disk (Variable Density Poisson disk) \cite{DWORK2021186}\cite{3d_poisson}. Despite these advancements, these conventional undersampling strategies are largely static. They rely on predefined geometric heuristics rather than the specific anatomical information underlying them. Consequently, achieving high spatial resolution still requires a relatively high sampling density to achieve adequate fidelity with acceptable Signal-to-Noise Ratio (SNR) under noisy conditions. This highlights that existing methods do not guarantee an optimal distribution of k-space samples that captures the maximum information per acquired point.

Here we present a new adaptive sampling technique for sampling the most optimal distribution of k-space coordinates in real time using an adaptive optimization algorithm utilizing state-of-the-art adiabatic quantum algorithms \cite{farhi2000quantumcomputationadiabaticevolution}. While work has been done on applying quantum algorithms for enhancing compressed sensing methods \cite{chevalier2024qaoa}, utilization of quantum algorithms directly in sampling in medical images has not been done yet. In this study, we formulate the selection of optimal k-space coordinates as a combinatorial optimization problem, drawing an analogy of atomic potential energy landscapes to design an Ising model Hamiltonian. The objective is to minimize the total Ising energy of the system, where "attraction" terms prioritize data points with high information density and "repulsion" terms enforce the incoherence required for CS. We further incorporate a Markov chain to track the previously acquired point allowing the algorithm to determine the next "best" coordinate based on the landscape information. The resulting non-convex optimization problem is solved using Quantum Annealing (QA), enabling the identification of near-global minima that classical greedy algorithms often overlook. 

We validated the proposed framework against standard undersampling trajectories, specifically Pseudo-Radial\cite{pseudo_radial}, Pseudo-Spiral\cite{pseudo_spiral}\footnote{Here in the paper the pseudo-radial and pseudo-spiral are also referred as radial and spiral respectively.}, Gaussian Random, and Variable-Density (VD) Poisson Disk sampling. For image reconstruction, we employed state-of-the-art compressed sensing architectures, including SENSE-TV \cite{sense_tv} optimized via the Alternating Direction Method of Multipliers (ADMM). The performance was rigorously evaluated across both single-coil and multi-coil (parallel imaging) k-space acquisitions using a comprehensive set of quantitative metrics: Peak Signal-to-Noise Ratio (PSNR), Structural Similarity Index (SSIM), High-Frequency Error Norm (HFEN), and Normalized Mean Square Error (NMSE). To further validate the robustness of our sampling algorithm, we tested it across different sampling rates and under simulated noisy conditions. The results indicate that Quantum Adaptive Sampling (QAS) maintains superior reconstruction fidelity even at aggressive, lower sampling rates where traditional methods typically suffer from significant aliasing. Furthermore, the QAS approach demonstrated consistent performance in noisy environments, suggesting that the quantum-optimized distribution is inherently more resilient to stochastic signal degradation.

Thus, the results demonstrate that, across all evaluated parameters and imaging conditions, our quantum adaptive sampling algorithm consistently outperforms traditional undersampling techniques. These findings suggest that quantum-driven adaptive masks provide superior reconstruction fidelity and lower error profiles in the MR images.

\section{Theory}

Traditionally, MRI sampling is done using a known trajectory (Spiral, Radial, etc.) \cite{eng2022goldenangle} or through known distributions. While these trajectories are effective for achieving basic incoherence, they lack the flexibility to adapt to the underlying signal's sparsity pattern. Conventional VD Poisson Disk sampling, for instance, relies on a static probability density function that does not account for the dynamic interactions between sampled points. This often results in a sub-optimal trade-off between the Signal-to-Noise Ratio (SNR) and the avoidance of coherent aliasing artifacts.

We model our technique by taking an analogy from the potential energy of an atomic system. The energy landscape of the k-space is partitioned into two competing terms: 1. Attraction term, which can be thought of as the nucleus attracting the electrons towards the center in an atom. This term acts as a potential well that prioritizes sampling in the k-space center. Since the bulk of the image's SNR and low-frequency structural information resides at the origin, this "nuclear" pull ensures that the most critical data points are densely sampled. 2. Repulsion term, similar to the electronic repulsion in a multi-electron atom. This term maintains an optimal spatial distance between sampled coordinates. Finally, the multi-electron atomic system is modeled as the Ising model. Let us see the detailed formulation of the multi-electron atomic potential in the following:
\subsection{Atomic Potential}
Let us consider a multi-electron atomic system with a nucleus at the center and $N$ electrons orbiting it. We define the distance from the $i^{th}$ electron to the nucleus as $r_i$ and the distance between the $i^{th}$ and $j^{th}$ electrons as $r_{ij}$ with $r_{ij} = |r_i - r_j|$. Therefore, the Coulomb's attractive potential is given by:
\begin{equation}
\label{eq:attractive_potential}
 V_{attractive} = -\sum_{i=1}^{N}\frac{Z}{r_i}
\end{equation}
and the multi-electron repulsive potential is given by:
\begin{equation}
\label{eq:repulsive_potential}
V_{repulsive} = \sum_{i=1}^{N}\sum_{j=1}^{N}\frac{e^2}{r_{ij}}
\end{equation}
with $Z$ being the atomic number (total number of protons in the nucleus) and $e$ being the electronic charge. 

Adding equations \eqref{eq:attractive_potential} and \eqref{eq:repulsive_potential}, we obtain the final Hamiltonian (total energy) for the atomic system.
\begin{equation}
    \label{eq:total_atomic_hamiltonian}
    H = -\sum_{i=1}^{N}\frac{Z}{r_i} + \sum_{i=1}^{N}\sum_{j=1}^{N}\frac{e^2}{r_{ij}}
\end{equation}

We use \eqref{eq:total_atomic_hamiltonian} as the Ising model Hamiltonian for our algorithm.

\subsection{Ising Model}
Metals have a spatially regular structure with atoms arranged in a lattice. Each atom possesses an angular momentum (spin), so it behaves like a bar magnet. Ferromagnetism arises when a collection of atomic spins align such that their associated magnetic moments all point in the same direction, and the spins behave like a big magnet with a net macroscopic magnetic moment. The statistical mechanics theory to describe ferromagnetism is called the Ising Model. The atomic spin can be either up or down. Based on this, the value of spin ($s$) is taken to be $+1(Up)$ or $-1(Down)$.

In a metal, the spins interact, flipping from up to down and vice versa. This spin-spin coupling and spin flipping under an external magnetic field constitute the total energy (Hamiltonian) of the Ising system, given by the following:
\begin{equation}
    \label{eq:Ising_hamiltonian}
    H_{ising} = \sum_{i=1}^{N}\sum_{j=1}^{N}J_{ij}s_is_j + \sum_{i=1}^{N}h_is_i
\end{equation}
where $N$ is the number of spins (atoms) in the metal, $J_{ij}$ is the spin-spin coupling energy, and $h_i$ is the spin interaction with the external magnetic field.

In Ising model formation, the Ising Hamiltonian is the objective function that needs to be minimized to obtain the optimal solution. We converted the problem of selecting the optimal k-space data into a combinatorial optimization problem to be solved by the Ising model.
\subsection{Adiabatic Quantum Computing}
Quantum mechanics was developed in the early $20^{th}$ century to study the physics of the atomic world. Using the power of quantum mechanics, quantum computation was developed to solve complex problems that classical computers cannot solve or will take millions of years to solve. Many quantum algorithms were developed in recent years, such as Grover's algorithm \cite{grover-1996}\cite{grover} for searching in an unsorted database, the Deutsch-Jozsa \cite{10.1098/rspa.1992.0167} algorithm for function determination exponentially faster than its classical counterparts, Shor's algorithm \cite{shorsalgorithm} to factor large integers in polynomial time and many more. 

A significant advancement in quantum computing for solving NP-hard combinatorial optimization problems is Adiabatic Quantum Computing (AQC). AQC utilizes the Adiabatic Theorem of quantum mechanics, which states that a quantum system will remain in its instantaneous ground state if the parameters of its Hamiltonian are varied sufficiently slowly.

The optimization problem is encoded into a \textit{problem Hamiltonian}($H_p$), whose ground state represents the global minimum of the objective function. The process begins with the system in the known state of the \textit{initial Hamiltonian}($H_0$). The system then undergoes a time-dependent evolution according to:
\begin{equation}
  H(t) = A(t)H_0 + B(t)H_p  
\end{equation}
where $t$ represents time, and $A(t)$ and $B(t)$ are annealing schedules that transition the system from $H_0$ to $H_p$ over a duration $T$. If the Hamiltonian is varied sufficiently slowly over an evolution time T, the system eventually terminates in the ground state of $H_p$.
\subsection{Quantum Annealing}
Quantum annealing (QA) \cite{quantum_annealing_implementation} is a physical realization of adiabatic quantum computing formulated to primarily solve combinatorial optimization problems. The quantum annealing algorithm is the quantum variant of simulated annealing \cite{simulated_annealing}, which is itself inspired by the metallurgical process of the same name. However, QA can be viewed as a metaheuristic that navigates complex energy landscapes by leveraging quantum tunneling to escape local minima that would trap classical algorithms like Simulated Annealing.

The physical implementation of QA typically maps the optimization problem onto a transverse-field Ising model \cite{qa}. The time-dependent Hamiltonian is expressed as:
\begin{equation}\hat{H}(s) = -\frac{A(s)}{2} \left( \sum_{i} \hat{\sigma}_{i}^{x} \right) + \frac{B(s)}{2} \left( \sum_{i} h_{i} \hat{\sigma}_{i}^{z} + \sum_{i<j} J_{ij} \hat{\sigma}_{i}^{z} \hat{\sigma}_{j}^{z} \right)\end{equation}
where $s$ is the normalized annealing parameter. The first term represents the tunneling Hamiltonian, which introduces quantum uncertainty by allowing spins to exist in a superposition of states. The second term is the problem Hamiltonian, where $h_i$ (local biases) and $J_{ij}$ (coupling strengths) encode the specific constraints of our k-space sampling mask.As the anneal progresses, the strength of the transverse field $A(s)$ is decreased to zero, while the problem Hamiltonian $B(s)$ is increased. If the process is sufficiently slow and the system remains relatively isolated from environmental decoherence, the qubits settle into a classical configuration ($+1$ or $-1$) that minimizes the Ising energy.

Quantum annealing can be used to solve problems modeled as Quadratic Unconstrained Binary Optimization (QUBO) or Ising problems.

\subsection{Parallel Tempering for Energy Optimization}
\label{sec:parallel_tempering}

Parallel tempering (PT), also known as replica exchange Monte Carlo, is a stochastic optimization technique designed to overcome the trapping of local search algorithms in metastable states of rugged energy landscapes~\cite{earl2005parallel, swendsen1986replica}. The method simultaneously simulates $N$ replicas of a system, each evolving at a distinct temperature $T_i$, where $i \in \{1, 2, \dots, N\}$ and $T_1 < T_2 < \cdots < T_N$. Each replica $i$ samples configurations from the Boltzmann distribution

\begin{equation}
    P_i(\mathbf{x}) = \frac{1}{Z_i} \exp\left(-\frac{E(\mathbf{x})}{k_B T_i}\right),
    \label{eq:boltzmann}
\end{equation}

where $E(\mathbf{x})$ is the energy (or cost/objective function) of configuration $\mathbf{x}$, $k_B$ is the Boltzmann constant, and $Z_i = \sum_{\mathbf{x}} \exp(-E(\mathbf{x})/k_B T_i)$ is the partition function at temperature $T_i$.

At high temperatures, replicas can traverse energy barriers freely, while low-temperature replicas refine solutions near local minima. Periodically, adjacent replicas $i$ and $i+1$ attempt to exchange their configurations $\mathbf{x}_i$ and $\mathbf{x}_{i+1}$. This swap is accepted according to the Metropolis criterion, with acceptance probability

\begin{equation}
    P_{\text{acc}}(i \leftrightarrow i+1) = \min\left\{1, \exp\left[\left(\frac{1}{k_B T_i} - \frac{1}{k_B T_{i+1}}\right)\left(E(\mathbf{x}_i) - E(\mathbf{x}_{i+1})\right)\right]\right\}.
    \label{eq:swap_acceptance}
\end{equation}

This exchange step satisfies detailed balance for the extended ensemble over all replicas,

\begin{equation}
    P(\mathbf{x}_1, \dots, \mathbf{x}_N) = \prod_{i=1}^{N} \frac{1}{Z_i} \exp\left(-\frac{E(\mathbf{x}_i)}{k_B T_i}\right),
    \label{eq:joint_distribution}
\end{equation}

ensuring that the overall Markov chain converges to the correct joint stationary distribution while allowing configurations to diffuse between temperature levels. As $t \to \infty$, the low-temperature replica $T_1$ approximates the global minimum of $E(\mathbf{x})$,

\begin{equation}
    \mathbf{x}^{*} = \arg\min_{\mathbf{x}} E(\mathbf{x}) \approx \lim_{t \to \infty} \mathbf{x}_1(t).
    \label{eq:global_min}
\end{equation}

The efficiency of PT depends critically on the choice of temperature ladder $\{T_i\}$, since the swap acceptance rate in Eq.~\eqref{eq:swap_acceptance} must remain sufficiently high (typically 20--40\%) to allow effective mixing between replicas~\cite{rozada2019effects}. A common heuristic is a geometric temperature schedule,

\begin{equation}
    T_i = T_1 \left(\frac{T_N}{T_1}\right)^{\frac{i-1}{N-1}}, \quad i = 1, \dots, N
    \label{eq:geometric_schedule}
\end{equation}

which balances exploration at high temperatures against exploitation at low temperatures, making PT particularly effective for combinatorial and continuous optimization problems with multimodal energy landscapes~\cite{earl2005parallel}.
\subsection{Mathematical Formulation}
Our proposed framework maps the MRI k-space grid to a system of interacting particles to leverage the energy minimization capabilities of quantum optimization. The acquisition process is partitioned into discrete temporal steps; in each step, a batch of $M$ k-space lines is selected from a larger candidate pool of $N$ randomly proposed coordinates. Thus, we convert each step into an optimization process for selecting the optimal coordinates. Each step following the initial acquisition is governed by a Markovian process, where the selection of the current batch is dynamically steered by the signal power measured in the preceding step. This feedback loop allows the Ising Hamiltonian to adapt its attraction landscape in real time, ensuring that sampling density is concentrated in high-information regions as the scan progresses.

We define the sampling decision as a binary configuration where each k-space coordinate $i$ is assigned a spin variable $s_i \in \{0, 1\}$, with $s_i=1$ indicating a selected line. The optimal sampling mask is found by minimizing the Ising Hamiltonian:
\begin{equation}
    \label{ising_energy}
    H_{ising} = \sum_{i<j} J_{ij} s_i s_j - \sum_i h_i s_i
\end{equation}
where $J_{ij}$ represents the interaction term and $h_i$ represents the attraction term.
 
\subsection{\textbf{1. Interaction Term ($J_{ij}$)}}

The interaction term is formulated as the electron-electron repulsion potential in the atomic model. To enforce spatial incoherence, we use an anisotropic distance $d_{ij}$ to account for separation along both the phase-encode ($y$) and partition ($z$) directions:
\begin{equation}
    \label{anisotropic_distance}
    d_{ij}^2 = (y_i - y_j)^2 + \alpha(z_i - z_j)^2
\end{equation}
The repulsion term is then formulated as a softened power law based on the equation \ref{eq:repulsive_potential}.
\begin{equation}
    \label{interaction_matrix}
    J_{ij} = \frac{\lambda}{(d_{ij}^2 + \epsilon)^{\gamma}}
\end{equation}
where $\epsilon$ is a small constant to introduce some offset, $\lambda$ is the energy scaling factor.

We introduce a softened distance $\tilde d_{ij}$ from the anisotropic distance in equation \ref{anisotropic_distance}:

\begin{equation}
\tilde{d}_{ij}^2 = d_{ij}^2 + 1.
\end{equation}

The base interaction strength is:
\begin{equation}
J_{ij}^{\text{base}} =
\frac{2}{\text{batch\_size}} \cdot
\left( \tilde{d}_{ij}^2 \right)^{-\gamma/2}.
\end{equation}

Finally, we obtain the interaction as:
\begin{equation}
J_{ij} =
\begin{cases}
0, & i = j, \\
J_{ij}^{\text{base}}, & \text{otherwise}.
\end{cases}
\end{equation}

\subsection{\textbf{2. Attraction Term ($h_i$)}}

The attraction term steers sampling toward the k-space center, while dynamically adapting to real-time signal power, similar to the central nuclear attraction in atomic systems. We divide the attraction term into three terms: a) \textit{static field} for central attraction, b) \textit{adaptive field} for dynamically steering sampling toward regions of high measured signal power via log-compression and spatial diffusion, and c) \textit{exploration bonus} for promoting frontier expansion and preventing overly clustered sampling in the k-space periphery. While keeping the inverse relation with distance as in equation \ref{eq:attractive_potential}, we model the static field $h_i^{static}$ for the $i^{th}$ line as:
\begin{equation}
    \label{h_static}
    h_i^{static} = exp(-\frac{r_i^2}{2\sigma^2})
\end{equation}
to ensure that the decay is faster with spatial distance.

We calculate $r_i$ and $\sigma$ as:
\begin{equation}
    \label{r_i}
    r_i^2 = (y-C)^2 + (z-C)^2
\end{equation}
where, $C = N/2$ as k-space center with $N$ being the matrix size.
\begin{equation}
    \sigma = N/(2c_b)
\end{equation}
where $c_b$ is the center bias.

We further normalize the field as follows to get the final $h_{static}$:
\begin{equation}
    \label{final_h_static}
    h_i^{static}\leftarrow \frac{2}{M}\frac{h_i^{static}}{max(h^{static})}
\end{equation}

Next, for the adaptive field, we first calculate the L2 norm of the previously acquired k-space data to obtain the signal ($s_i$) as $s_i = ||k_i||$. Then we use logarithmic suppression to reduce the dynamic range and prevent the dominance of extremely bright regions as $a_i = log(1+s_i)$. In addition, we add a Gaussian smoothing kernel to allow neighborhood exploration through $\tilde a_i = G_{\sigma_g}*a_i$. Finally, we normalize the value to obtain the adaptive field $h_i^{adaptive}$ as:
 
\begin{equation}
    \label{h_adaptive}
    h_i^{adaptive} \leftarrow \frac{\tilde a_i}{max(\tilde a)}
\end{equation}
Now, to prevent the adaptive field from overwhelming
spatial repulsion, we apply a non-linear suppression using the hyperbolic function \textit{i.e}
\begin{equation}
    \label{final_h_adaptive}
    h_i^{adaptive} \leftarrow tanh(\beta h_i^{adaptive})
\end{equation}

Third, the exploration bonus $b_i$ is calculated as follows:
\begin{equation}
    \label{exploration_bonus}
    b_i = \eta \frac{\text{Number of Un-sampled Neighbors}}{\text{Number of Neighbors}}
\end{equation}
where $\eta$ is the exploration strength.

Therefore, from equations \ref{final_h_static}, \ref{final_h_adaptive}, and \ref{exploration_bonus}, the attraction term becomes:
\begin{equation}
    \label{final_h}
    h_i=h_i^{static} + w_ah_i^{adaptive} + b_i
\end{equation}
where $w_a$ is the adaptive weight controlling data influence.

Finally, we add equations \ref{interaction_matrix} and \ref{final_h} for all candidate lines to obtain our Ising Hamiltonian. We also add a penalty term to ensure that the QPU or simulated solvers select exactly $M$ lines per batch.
\begin{equation}
    H_{total}  = \sum_{i<j}J_{ij} s_i s_j - \sum_i h_i s_i + P\left(\sum_i s_i - M\right)^2
\end{equation}

Once a coordinate is sampled, it is masked by setting $h_i = -\infty$, thereby ensuring the iterative batch selection process never re-selects previously sampled data points.

The workflow in Algorithm~1 summarizes the outer acquisition loop. To make the inner steps explicit, Algorithms~2--4 describe how the adaptive field is updated from the previous batch, how the interaction matrix is assembled for a candidate pool \ref{pool}, and how the constrained batch-selection problem is formed before solving.

\begin{algorithm}[H]
\SetAlgoLined
\DontPrintSemicolon

\KwIn{Previously measured batch data $k$, candidate coordinates $\mathcal{C}$, center bias $c_b$, smoothing width $\sigma_g$, adaptive weight $w_a$, suppression factor $\beta$, exploration strength $\eta$}
\KwOut{Attraction field $\{h_i\}_{i \in \mathcal{C}}$}

Compute the static center-biased field $h_i^{static}$ for all $i \in \mathcal{C}$ using equations \ref{h_static}--\ref{final_h_static}\;

\ForEach{previously sampled location $i$}{
    Compute signal magnitude $s_i \leftarrow \|k_i\|_2$\;
    Apply log compression $a_i \leftarrow \log(1+s_i)$\;
}

Diffuse the compressed signal with a Gaussian kernel: $\tilde{a} \leftarrow G_{\sigma_g} * a$\;
Normalize the diffused map and apply suppression: $h_i^{adaptive} \leftarrow \tanh\!\left(\beta \, \tilde{a}_i / \max(\tilde{a})\right)$\;

\ForEach{candidate coordinate $i \in \mathcal{C}$}{
    Compute the frontier bonus $b_i$ using equation \ref{exploration_bonus}\;
    Set $h_i \leftarrow h_i^{static} + w_a h_i^{adaptive} + b_i$\;
}

\Return{$\{h_i\}_{i \in \mathcal{C}}$}\;
\caption{Adaptive attraction-field update}
\end{algorithm}

\begin{algorithm}[H]
\SetAlgoLined
\DontPrintSemicolon

\KwIn{Candidate coordinates $\mathcal{C}$, anisotropy $\alpha$, decay $\gamma$, batch size $M$}
\KwOut{Interaction matrix $J$}

\ForEach{unordered pair $(i,j)$ with $i,j \in \mathcal{C}$ and $i \neq j$}{
    Compute anisotropic distance $d_{ij}^2 \leftarrow (y_i-y_j)^2 + \alpha (z_i-z_j)^2$\;
    Form softened distance $\tilde d_{ij}^2 \leftarrow d_{ij}^2 + 1$\;
    Compute base interaction $J_{ij}^{base} \leftarrow \frac{2}{M}\left(\tilde d_{ij}^2\right)^{-\gamma/2}$\;
    Set $J_{ij} \leftarrow J_{ij}^{base}$\;
}

Set $J_{ii} \leftarrow 0$ for all $i \in \mathcal{C}$ and symmetrize $J$\;
\Return{$J$}\;
\caption{Pairwise interaction-matrix construction}
\end{algorithm}

\begin{algorithm}[H]
\SetAlgoLined
\DontPrintSemicolon

\KwIn{Candidate pool $\mathcal{C}$, interaction matrix $J$, attraction field $h$, batch size $M$, penalty $P$}
\KwOut{Selected batch $\mathcal{B}$}

Assign one binary variable $s_i \in \{0,1\}$ to each $i \in \mathcal{C}$\;
Form the batch objective
\[
H_{total} = \sum_{i<j \in \mathcal{C}} J_{ij}s_is_j - \sum_{i \in \mathcal{C}} h_i s_i + P\left(\sum_{i \in \mathcal{C}} s_i - M\right)^2
\]
Solve the constrained Ising/QUBO problem with the selected backend\;
Collect the coordinates with $s_i = 1$ into $\mathcal{B}$\;
Mask $\mathcal{B}$ from future candidate pools so they are not re-selected\;

\Return{$\mathcal{B}$}\;
\caption{Constrained batch selection from a candidate pool \ref{pool}}
\end{algorithm}

 We also add a penalty term to ensure that the QPU or simulated solvers select exactly $M$ lines per batch.
\begin{equation}
    H_{total}  = \sum_{i<j}J_{ij} s_i s_j - \sum_i h_i s_i + P\left(\sum_i s_i - M\right)^2
\end{equation}

Once a coordinate is sampled, it is masked by setting $h_i = -\infty$, thereby ensuring the iterative batch selection process never re-selects previously sampled data points.

\begin{algorithm}[H]
\SetAlgoLined
\DontPrintSemicolon

\KwIn{Target sampling rate $R_t$, batch size $M$, candidate pool size $N$, penalty $P$}
\KwOut{Optimized k-space sampling mask $S$}

\textbf{Initialization:}\;
Randomly select $M$ k-space lines to form the initial mask $S$\;
Compute initial system energy\;

\While{Sampling rate $<$ $R_t$}{
    
    \textbf{Candidate Generation:}\;
    Generate a candidate pool of $N$ unsampled k-space lines\;
    
    \textbf{Attraction Field Update:}\;
    \eIf{Total samples $<$ threshold ($\tau$)}{
        Update attraction field with constant center bias\;
    }{
        Update attraction field with decreasing center bias\;
    }
    
    \textbf{Interaction Estimation:}\;
    Compute pairwise interaction matrix\;
    
    \textbf{Quantum Optimization:}\;
    Formulate the sampling selection as an Ising optimization problem\;
    Solve on the QPU to select the optimal batch of $M$ lines\;
    
    \textbf{Mask Update and Feedback:}\;
    Acquire selected k-space lines and update mask $S$\;
    Update adaptive attraction field using measured signal power\;
    Disable already sampled locations from further selection\;
}

\Return{$S$}\;

\caption{Quantum Adaptive Compressed Sensing (QAS) Workflow}
\end{algorithm}

\section{Methods}

We evaluated QAS using the enhanced-energy formulation described in equation~\ref{final_h_adaptive} and solved the corresponding sampling problem with parallel tempering. The study was designed as an 8-coil fair-budget comparison in which QAS was measured against variable-density Poisson disk (VDP), Cartesian variable-density, radial, spiral, and uniform-random sampling. All experiments were performed on $128^3$ and $256^3$ phantoms using the same total sampling budget at each acceleration factor. The phantom used for the simulation is a digital image of brain from MPRAGE protocol and then we simulated coil sensitivity matrix to multiply with the image and doing an Fast Fourier Transform (FFT) gave us full sampled ground truth to work with.
When an external CSM file is not supplied, we generate an analytical
birdcage-like CSM for simulation.  Coil centers are placed uniformly on a
ring in the $xy$-plane, and the resulting complex sensitivities are
normalized independently at every voxel.

\begin{algorithm}[H]
\SetAlgoLined
\DontPrintSemicolon

\KwIn{Spatial size $N$, number of coils $C$, ring radius $r_c=0.8$, sensitivity width $w=0.6$, decay exponent $p=2$}
\KwOut{Normalized complex CSM $S_c[z,y,x]$ for $c=0,\ldots,C-1$}

\For{$c \leftarrow 0$ \KwTo $C-1$}{
    Set coil angle $\phi_c \leftarrow 2\pi c/C$ and position
    $\mathbf{p}_c \leftarrow (r_c\cos\phi_c, r_c\sin\phi_c, 0)$\;
    Set inward unit direction
    $\mathbf{u}_c \leftarrow (-\cos\phi_c,-\sin\phi_c,0)$\;
}

\ForEach{voxel $(z,y,x)$ in the $N\times N\times N$ grid}{
    Map its indices to normalized coordinates
    $(n_x,n_y,n_z) \leftarrow (2x/(N-1)-1,\,2y/(N-1)-1,\,2z/(N-1)-1)$\;
    \For{$c \leftarrow 0$ \KwTo $C-1$}{
        Compute displacement $\mathbf{d}_{c} \leftarrow (n_x,n_y,n_z)-\mathbf{p}_c$
        and $\delta_c \leftarrow \sqrt{\|\mathbf{d}_c\|_2^2+10^{-6}}$\;
        Compute radial falloff
        $m_c \leftarrow \left[1+(\delta_c/w)^p\right]^{-1}$\;
        Compute directional weight
        $q_c \leftarrow 0.3+0.7\left[(1+(\mathbf{d}_c\cdot\mathbf{u}_c)/(\delta_c+10^{-6}))/2\right]^{1/2}$\;
        Set phase $\theta_c \leftarrow (\pi/2)(\cos\phi_c\,n_x+\sin\phi_c\,n_y)$\;
        Set unnormalized sensitivity
        $S_c[z,y,x] \leftarrow m_c q_c\exp(\mathrm{i}\theta_c)$\;
    }
    Compute $a \leftarrow \left(\sum_{c=0}^{C-1}|S_c[z,y,x]|^2+10^{-10}\right)^{-1/2}$\;
    \For{$c \leftarrow 0$ \KwTo $C-1$}{
        Normalize $S_c[z,y,x] \leftarrow a\,S_c[z,y,x]$\;
    }
}

\Return{$\{S_c\}_{c=0}^{C-1}$}\;
\caption{Analytical birdcage-like coil-sensitivity-map generation}
\label{alg:analytical_csm}
\end{algorithm}

We used the anisotropic distance in equation \ref{anisotropic_distance} between the phase-encode and partition directions with $\alpha = 2$ and $\gamma=2.5$. Additionally, we use the adaptive field with $tanh$ suppression as in \ref{final_h_adaptive} with $\beta = 2$. Grid search \cite{grid_search_1} \cite{grid_search_2} algorithm was applied to determine the suitable parameters. 

We use thresholding here to prevent the solver from over-exploiting the high-energy central k-space region. Therefore, the algorithm uses a linearly decreasing center bias for the attractive field. For the experiments, a threshold ($\tau$) of $85\%$ of the total number of k-space lines to be sampled. So, $\tau=0.85 \times R_t \times (N)^2$ for our specific case.

We validated our quantum adaptive sampling algorithm against spiral\cite{pseudo_spiral}, radial \cite{eng2022goldenangle}, random, and variable-density Poisson-disk \cite{DWORK2021186} \cite{3d_poisson}, which are standard undersampling techniques. The under-sampled k-space data generated by masks produced by these methods were reconstructed using the primary state-of-the-art architectures: SENSE-TV ADMM \cite{sense_tv}.

For each phantom size, we generated undersampling masks for two acceleration settings: $R=5$ (20\% sampling) and $R=10$ (10\% sampling). Further, noise robustness was assessed by injecting per-batch complex Gaussian noise in k-space and repeating the reconstructions at clean, 40, 30, 25, and 20\,dB input SNR. The noise was injected in four stages: first, the signal power of the k-space data was computed; next, the target SNR was converted from dB to a linear scale as 

\begin{equation}
    SNR_{linear} = 10^{\frac{SNR_{db}}{10}}
    \label{input_snr_calculation}
\end{equation}

which was used to determine the corresponding noise variance; Gaussian noise with this variance was then generated; and finally, complex noise was added to both real and imaginary components with equal variance.

All the methods were compared under a strict fair budget constraint, and it was further tested that any sampling rate above $25\%$ yielded similar results for both QAS and VDP, confirming that our adaptive algorithm outperforms other methods at lower sampling rates and has diminishing results at higher sample amounts as the information acquired becomes similar.

For SENSE-TV reconstruction, the image estimate was obtained by solving

\begin{equation}
\hat{x}=\arg\min_x\frac{1}{2}\left\|P_{\Omega} F S x - y_{\Omega}\right\|_2^2+\lambda_{\mathrm{TV}}\|\nabla x\|_1,
\label{eq:sense_tv}
\end{equation}

Here, $x$ denotes the reconstructed image, $S$ is the coil sensitivity operator, $F$ is the Fourier encoding operator, $P_{\Omega}$ is the under sampling mask that selects the acquired k-space samples, $y_{\Omega}$ represents the acquired under-sampled multi-coil k-space measurements, $\nabla$ denotes the spatial finite-difference gradient operator, and $||\nabla x||_{1}$ is the isotropic total variation (TV) regularization term. The TV regularization parameter ($\lambda_{TV}$) was set to $5\times10^{-4}$. The value of $\lambda_{TV}$ is kept the same across all the methods.

NMSE was computed as
\begin{equation}
\mathrm{NMSE}=\frac{\|\hat{x}-x_{\mathrm{ref}}\|_2^2}{\|x_{\mathrm{ref}}\|_2^2}.
\end{equation}

HFEN\cite{hfen} was computed as
\begin{equation}
\mathrm{HFEN}=\frac{\|\mathrm{LoG}_{\sigma}(\hat{x})-\mathrm{LoG}_{\sigma}(x_{\mathrm{ref}})\|_2}{\|\mathrm{LoG}_{\sigma}(x_{\mathrm{ref}})\|_2},
\end{equation}
where $\mathrm{LoG}_{\sigma}$ denotes a Laplacian-of-Gaussian high-pass filter.

PSNR was calculated as
\begin{equation}
\mathrm{PSNR}
=
10 \log_{10}
\left(
\frac{(\mathrm{DR})^2}{\mathrm{MSE}}
\right),
\end{equation}

where
\begin{equation}
\mathrm{MSE}
=
\frac{1}{N}
\left\|
\hat{x}-x_{\mathrm{ref}}
\right\|_2^2,
\end{equation}

and \(\mathrm{DR}\) denotes the dynamic range of the reference image.

And finally, SSIM is computed as

\begin{equation}
\mathrm{SSIM}(x_{\mathrm{ref}},\hat{x})
=
\frac{
(2\mu_{x_{\mathrm{ref}}}\mu_{\hat{x}} + C_1)
(2\sigma_{x_{\mathrm{ref}}\hat{x}} + C_2)
}{
(\mu_{x_{\mathrm{ref}}}^2 + \mu_{\hat{x}}^2 + C_1)
(\sigma_{x_{\mathrm{ref}}}^2 + \sigma_{\hat{x}}^2 + C_2)
},
\end{equation}

where \(\mu_{x_{\mathrm{ref}}}\) and \(\mu_{\hat{x}}\) are the local mean intensities,
\(\sigma_{x_{\mathrm{ref}}}^2\) and \(\sigma_{\hat{x}}^2\) are the local variances,
and \(\sigma_{x_{\mathrm{ref}}\hat{x}}\) is the local covariance between the reference and reconstructed images.
The constants \(C_1\) and \(C_2\) are included to stabilize the division for small denominator values.

With a unique k-space sample each time, the reconstruction quality was quantified with Peak Signal-to-Noise Ratio (PSNR) and Structural Similarity Index (SSIM) as primary fidelity measures. Normalized Mean Square Error (NMSE) and High-Frequency Error Norm (HFEN) were also evaluated to characterize overall reconstruction error and preservation of fine structure. Results are reported across both matrix sizes, both reconstruction methods, and both acceleration factors.

We performed the experiments using the classical annealing technique Parallel Tempering (PT), which extends simulated annealing by evolving multiple replicas of the system at different temperatures and periodically exchanging configurations between them. This enables more efficient exploration of the energy landscape and reduces the likelihood of trapping in local minima, providing a strong classical baseline for comparison with our proposed sampling strategy. 

\subsubsection{Pooling \& D'Wave Implementation}
\label{pool}
Pooling is essentially selecting a group of k-space lines from the entire k-space at random, mainly due to resource constraints for the quantum annealer to reduce the overall search space per batch that is being optimized. For example, from a k-space with $\textbf{T}$ k-space lines, only $\zeta$ lines are selected, where $\zeta \leq \textbf{T}$. Thus, the candidate pool is the full k-space for simulations. During the run on the D'Wave due to the size of the Hamiltonian to be solved, it was hitting the computational limits of the D'Wave machine, so to solve the issue, we decided to subsample the Hamiltonian by taking a sample out of it, so the values don't change, but the selected indices are mapped back to the original indices for the selection procedure.
We implemented the algorithm using D-Wave's Leap Hybrid Binary Quadratic Model (BQM) Solver ($hybrid\_binary\_quadratic\_model$ version2p, version 2.2). This cloud-based hybrid quantum-classical solver automatically decomposes large BQM instances and dynamically selects the most suitable available D-Wave Quantum Processing Unit (QPU)(D-Wave Quantum Inc, Boca Raton, Florida) for solving subproblems, while coordinating the optimization through proprietary classical heuristics. During our experiments, the available QPUs included Advantage2\_system1.11 based on the Zephyr (Z12) topology with 4,577 physical qubits and 41,515 couplers, Advantage\_system4.1 based on the Pegasus (P16) topology with 5,627 physical qubits and 40,279 couplers, and Advantage\_system6.4, also based on the Pegasus (P16) topology with 5,612 physical qubits and 40,088 couplers. The hybrid solver automatically utilizes the appropriate QPU depending on the problem characteristics, abstracting the embedding, decomposition, and quantum-classical orchestration from the user. Experiments were conducted on a $128\times128\times128$ Shepp-Logan phantom with a target sampling rate of 10\%, selecting a batch of 48 samples from a pool of 4096 given the limit of physical qubits. 
\section{Results}

The performance of the Quantum Adaptive Sampling (QAS) framework was rigorously evaluated against standard sampling methods \textit{viz.} Variable-Density Poisson (VDP) disk, pseudo-radial, pseudo-spiral, cartesian, and uniform random sampling across multiple acquisition scenarios. Findings demonstrate that QAS consistently achieves superior reconstruction fidelity, with the advantage becoming more pronounced at higher acceleration factors and under significant noise interference.

\subsection{Performance Comparison: Noiseless Configurations}

We employed Parallel Tempering as the annealing-based optimization solver for the proposed quantum-optimized sampling algorithm. All experiments employed an 8-coil receive configuration with SENSE-TV reconstruction to evaluate the proposed sampling masks under a parallel imaging framework.

In the noiseless setting, QAS achieved the highest reconstruction fidelity across both acceleration factors for the $256^3$ phantom. At $R=5$, corresponding to a $20\%$ sampling rate, QAS obtained a PSNR of $49.03$ dB, SSIM of $0.9962$, NMSE of $0.0030$, and HFEN of $0.0437$. These results outperform all competing sampling strategies, including VDP, Cartesian VD, radial, spiral, and uniform random sampling. Compared with VDP, the strongest baseline in this setting, QAS improved PSNR by $4.65$ dB while also reducing NMSE from $0.0087$ to $0.0030$ and HFEN from $0.0798$ to $0.0437$. This indicates that the proposed sampling strategy improves not only global voxel-wise fidelity, but also preservation of high-frequency image structure.

\begin{table}[H]
\centering
\begin{tabular}{c|l|cccc}
\hline
\textbf{Acceleration} & \textbf{Method} & \textbf{PSNR} $\uparrow$ & \textbf{SSIM} $\uparrow$ & \textbf{NMSE} $\downarrow$ & \textbf{HFEN} $\downarrow$ \\
\hline
\multirow{6}{*}{$R=5$}
& QAS & \textbf{49.03} & \textbf{0.9962} & \textbf{0.0030} & \textbf{0.0437} \\
& VDP & 44.38 & 0.9917 & 0.0087 & 0.0798 \\
& Cartesian VD & 30.45 & 0.8835 & 0.2161 & 0.4578 \\
& Radial & 37.51 & 0.9664 & 0.0425 & 0.1969 \\
& Spiral & 34.30 & 0.9473 & 0.0891 & 0.2398 \\
& Uniform Random & 34.32 & 0.9218 & 0.0886 & 0.3343 \\
\hline
\multirow{6}{*}{$R=10$}
& QAS & \textbf{36.66} & \textbf{0.9650} & \textbf{0.0517} & \textbf{0.2033} \\
& VDP & 34.22 & 0.9468 & 0.0907 & 0.2822 \\
& Cartesian VD & 25.87 & 0.7603 & 0.6204 & 0.8135 \\
& Radial & 28.49 & 0.8361 & 0.3394 & 0.5941 \\
& Spiral & 22.11 & 0.6545 & 1.4724 & 0.5370 \\
& Uniform Random & 28.31 & 0.6965 & 0.3534 & 0.8851 \\
\hline
\end{tabular}
\caption{Noiseless reconstruction performance for the $256^3$ phantom using 8-coil SENSE-TV reconstruction. The best result for each metric and acceleration factor is shown in bold.}
\label{tab:noiseless_sense_tv_n256}
\end{table}

The advantage of QAS persisted under more aggressive undersampling. At $R=10$, corresponding to a $10\%$ sampling rate, QAS again achieved the best performance across all metrics, with a PSNR of $36.66$ dB, SSIM of $0.9650$, NMSE of $0.0517$, and HFEN of $0.2033$. Relative to VDP, QAS provided a $2.44$ dB PSNR improvement and reduced NMSE by approximately $43\%$. The improvement in HFEN further suggests that QAS better preserves edge-like and high-spatial-frequency features, which are particularly vulnerable under severe undersampling. This is consistent with the qualitative evaluations, where QAS exhibits sharper boundaries, fewer structured artifacts, and lower residual error compared with the baseline sampling patterns as seen in Figure \ref{fig:recon_zoom_combined} where the top two performing methods are compared to the ground truth fully sampled noiseless reconstruction .

The comparison across sampling strategies also highlights the limitations of conventional heuristic designs. Cartesian VD sampling produced substantially lower PSNR and SSIM, indicating that its structured undersampling artifacts were not fully mitigated by SENSE-TV reconstruction. 
Radial and spiral sampling performed better than Cartesian VD in some cases, but they remained clearly below QAS, suggesting that trajectory-inspired coverage alone was insufficient to optimize reconstruction fidelity for this volumetric setting. 
Uniform random sampling also underperformed, emphasizing that incoherence by itself does not guarantee optimal reconstruction when the sampling distribution is not adapted to the image structure and reconstruction prior.

Similar trends were observed for the $128^3$ phantom, further confirming that the benefit of QAS is not limited to the higher-resolution $256^3$ setting. Table~\ref{tab:noiseless_sense_tv_n128} summarizes the noiseless reconstruction performance for the $128^3$ phantom under 8-coil SENSE-TV reconstruction. At $R=5$, corresponding to a $20\%$ sampling rate, QAS achieved the best performance across all reported metrics, with a PSNR of $37.47$ dB, SSIM of $0.9744$, NMSE of $0.0306$, and HFEN of $0.1556$. Compared with VDP, QAS improved PSNR by $3.85$ dB and increased SSIM by $0.0240$, while also substantially reducing NMSE and HFEN. This indicates that the proposed sampling strategy improves both global reconstruction fidelity and high-frequency structural preservation in the lower-resolution phantom.

\begin{table}[H]
\centering
\begin{tabular}{c|l|cccc}
\hline
\textbf{Acceleration} & \textbf{Method} & \textbf{PSNR} $\uparrow$ & \textbf{SSIM} $\uparrow$ & \textbf{NMSE} $\downarrow$ & \textbf{HFEN} $\downarrow$ \\
\hline
\multirow{6}{*}{$R=5$}
& QAS & \textbf{37.47} & \textbf{0.9744} & \textbf{0.0306} & \textbf{0.1556} \\
& VDP & 33.62 & 0.9504 & 0.0742 & 0.2578 \\
& Cartesian VD & 27.07 & 0.8259 & 0.3348 & 0.5158 \\
& Radial & 30.30 & 0.8872 & 0.1594 & 0.3677 \\
& Spiral & 30.09 & 0.9062 & 0.1672 & 0.3129 \\
& Uniform Random & 33.96 & 0.9289 & 0.0685 & 0.3018 \\
\hline
\multirow{6}{*}{$R=10$}
& QAS & \textbf{27.57} & \textbf{0.8542} & \textbf{0.2989} & \textbf{0.5245} \\
& VDP & 26.93 & 0.8280 & 0.3463 & 0.5591 \\
& Cartesian VD & 23.44 & 0.6789 & 0.7732 & 0.9134 \\
& Radial & 24.66 & 0.7342 & 0.5831 & 0.7235 \\
& Spiral & 23.09 & 0.6480 & 0.8375 & 0.7931 \\
& Uniform Random & 23.88 & 0.6485 & 0.6986 & 0.6872 \\
\hline
\end{tabular}
\caption{Noiseless reconstruction performance for the $128^3$ phantom using 8-coil SENSE-TV reconstruction. The best result for each metric and acceleration factor is shown in bold.}
\label{tab:noiseless_sense_tv_n128}
\end{table}

The advantage of QAS was preserved under stronger undersampling. At $R=10$, corresponding to a $10\%$ sampling rate, QAS again achieved the highest PSNR and SSIM, as well as the lowest NMSE and HFEN among all methods. Although the performance gap relative to VDP was smaller at $R=10$ than at $R=5$, QAS still provided a consistent improvement, increasing PSNR from $26.93$ dB to $27.57$ dB and SSIM from $0.8280$ to $0.8542$. This result is important because it shows that the proposed optimization does not merely improve reconstruction in moderately accelerated settings, but remains beneficial when the sampling budget is reduced to only $10\%$ of k-space.

Overall, the $128^3$ results support the same conclusion observed in the $256^3$ phantom: QAS consistently yields sampling masks that are better matched to the SENSE-TV reconstruction model than conventional heuristic sampling patterns. The improvement across PSNR, SSIM, NMSE, and HFEN suggests that QAS enhances both intensity fidelity and structural preservation, while reducing residual reconstruction error. Together, the noiseless experiments across $128^3$ and $256^3$ resolutions demonstrate that the proposed quantum-optimized sampling framework generalizes across phantom size and acceleration factor. 

For the 256³ experiments, the QAS Ising optimization was solved using parallel tempering in batches. For acceleration factor R = 5, the Ising solver required 274 batches per run, with total solver times ranging from 58.69 s to 70.47 s across SNR levels. This corresponds to an average parallel-tempering time of 214.2–257.2 ms\footnote{The algorithm is not GPU accelerated, as the simulation bottleneck is the reconstruction.} per batch. For R = 10, the solver required 137 batches per run, with total solver times ranging from 32.26 s to 47.34 s, corresponding to an average of 235.5–345.6 ms per batch. Across all 256³ experiments, the complete QAS Ising optimization accounted for only 8.34 min of runtime, compared with 8.51 h spent on image reconstruction.

\subsection{Noise Resilience and Real-Time Adaptive Acquisition}

We evaluated the noise robustness of the proposed QAS framework under SENSE-TV reconstruction using 8-coil configurations for both $256^3$ and $128^3$ phantoms. With SENSE-TV reconstruction, overall stability for all sampling strategies was achieved; however, QAS delivered the most robust performance across both resolutions and acceleration factors. For the $256^3$ phantom, QAS maintained consistently high reconstruction quality, achieving PSNR values up to $\sim 36.75$ dB at $R=10$, while VDP and radial showed greater degradation under moderate noise. 

\begin{figure}[H]
    \centering
    \includegraphics[width=\linewidth]{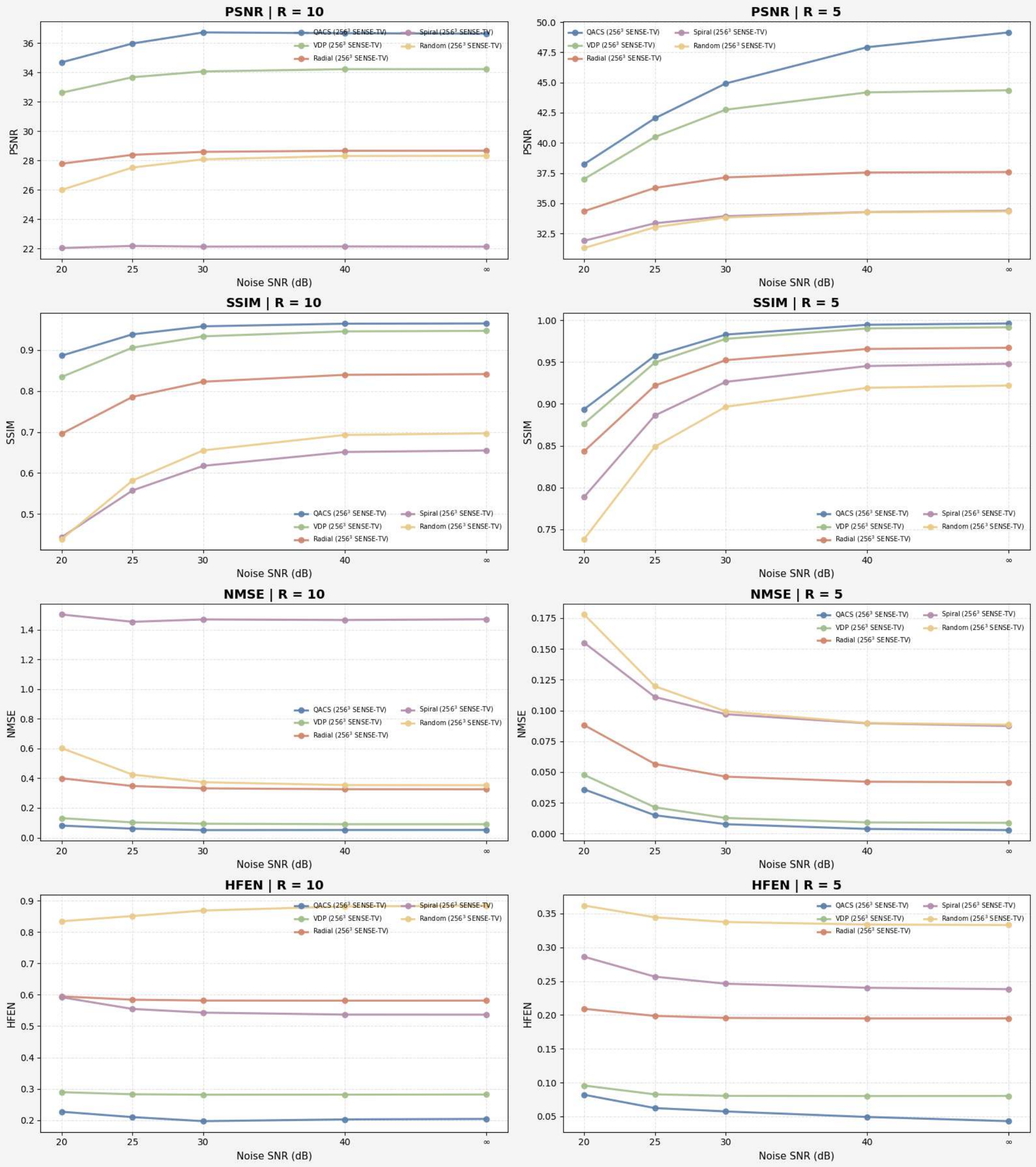}
    \caption{Comparison across Sampling Methods under Noisy Situations ($256^3$) at $R=5$ and $R=10$}
    \label{fig:256_noise_study}
\end{figure}

\begin{figure}[H]
    \centering
    \includegraphics[width=\linewidth]{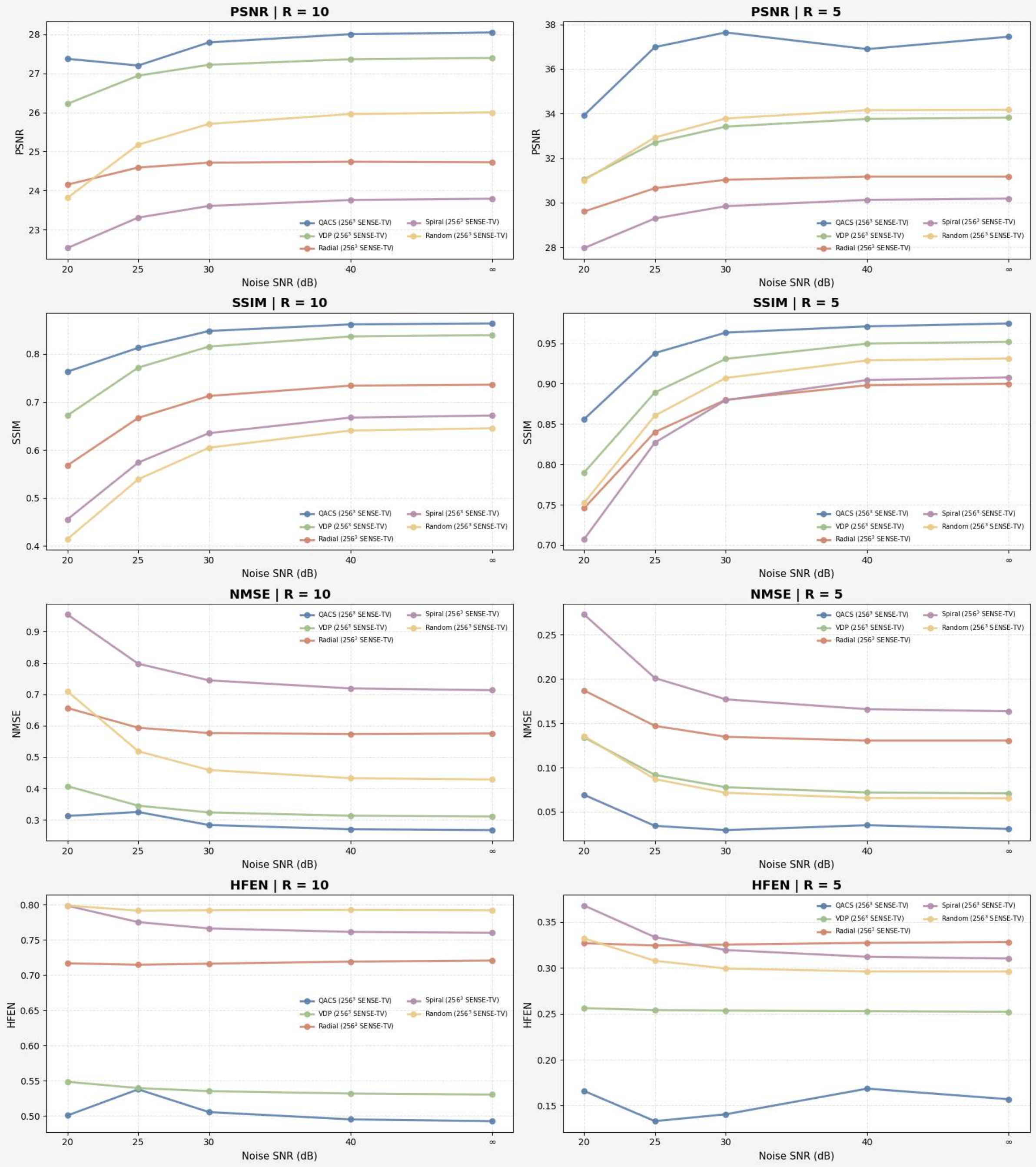}
    \caption{Comparison across Sampling Methods under Noisy Situations ($128^3$) at $R=5$ and $R=10$}
    \label{fig:128_noise_study}
\end{figure}

However, under the most severe simulated noise condition of 20 dB input SNR (using Equation \ref{input_snr_calculation}), reconstruction quality decreased for all sampling strategies. Although the performance differences narrowed in some settings, QAS retained a noticeable advantage over VDP at R=10, suggesting that adaptive sampling remains beneficial when aggressive undersampling is combined with substantial measurement noise.
At moderate noise levels (30dB of input SNR as per the Equation \ref{input_snr_calculation}), across both $R=5$ and $R=10$, QAS exhibited the flattest degradation curves in PSNR and SSIM, along with the lowest HFEN, highlighting superior edge preservation and resistance to noise-induced artifacts. Spiral and uniform random sampling remained the least robust, with a rapid loss of structural detail at lower SNR. Similar behavior was observed for the $128^3$ phantom, where QAS consistently preserved higher structural fidelity and lower reconstruction error compared to all baselines, including VDP. Overall, these results indicate that the quantum-adaptive sampling strategy not only remains robust under moderate noisy conditions but also complements advanced regularized reconstruction, enabling improved structural preservation and reduced error propagation across varying resolutions and sampling regimes. Furthermore, we observe convergence of all algorithms in the reconstructed image parameters at higher sampling rates, since most of the k-space is captured by all of them.
The variations of PSNR, SSIM, HFEN and NMSE across various noise levels are shown through Fig. \ref{fig:256_noise_study} to \ref{fig:128_noise_study}.

\begin{figure}[H]
    \centering
    \includegraphics[width=0.875\linewidth]{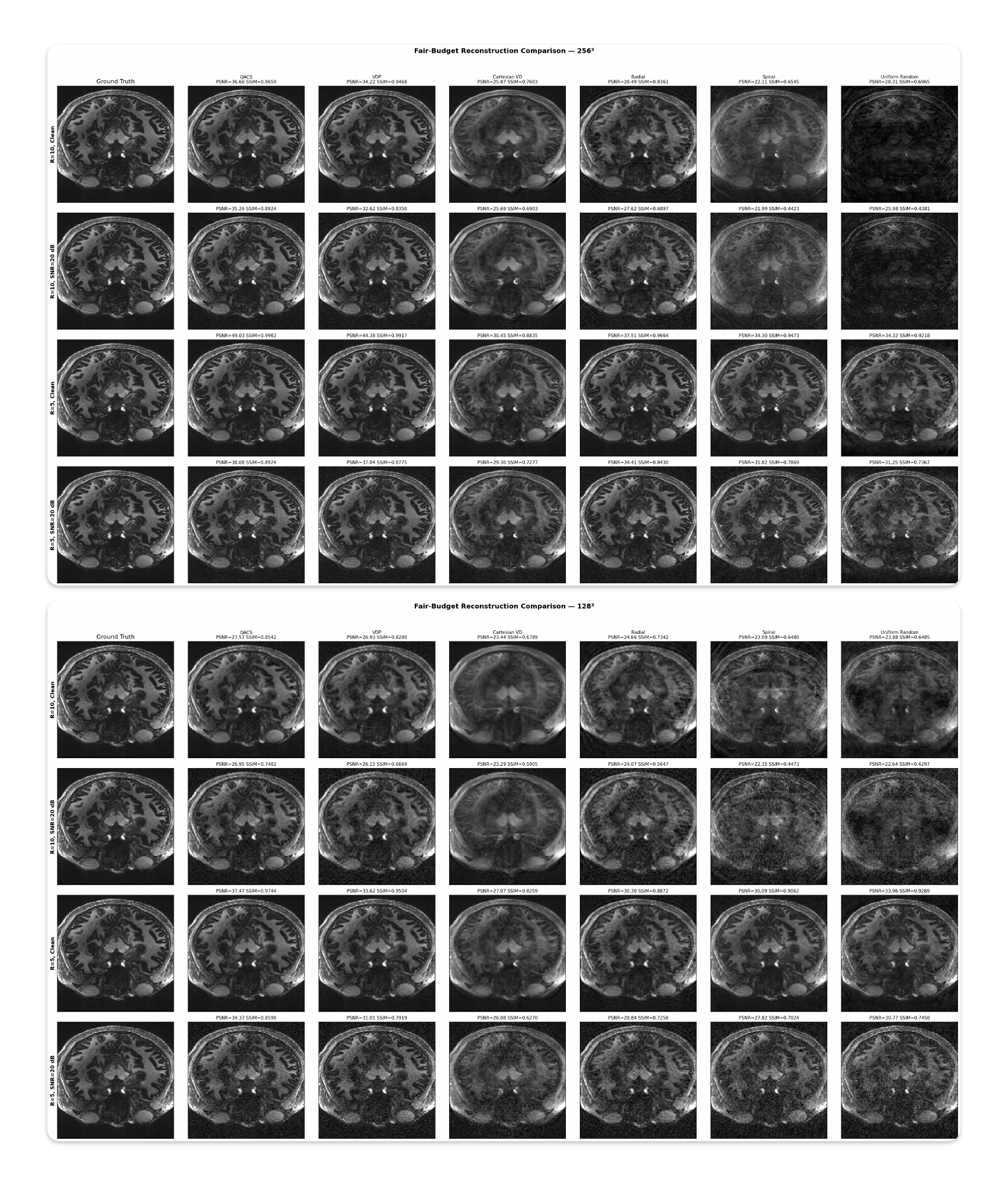}
    \caption{Qualitative comparison of reconstructed images for $128^3$ and $256^3$ phantoms under SENSE-TV reconstruction at $R=5$ and $R=10$ , evaluated across varying noise levels. QAS consistently preserves structural details and suppresses noise more effectively than conventional sampling strategies, with the performance gap widening at higher acceleration and lower SNR.}
    \label{fig:recon_combined}
\end{figure}

\begin{figure}[H]
    \centering
    \includegraphics[width=\linewidth]{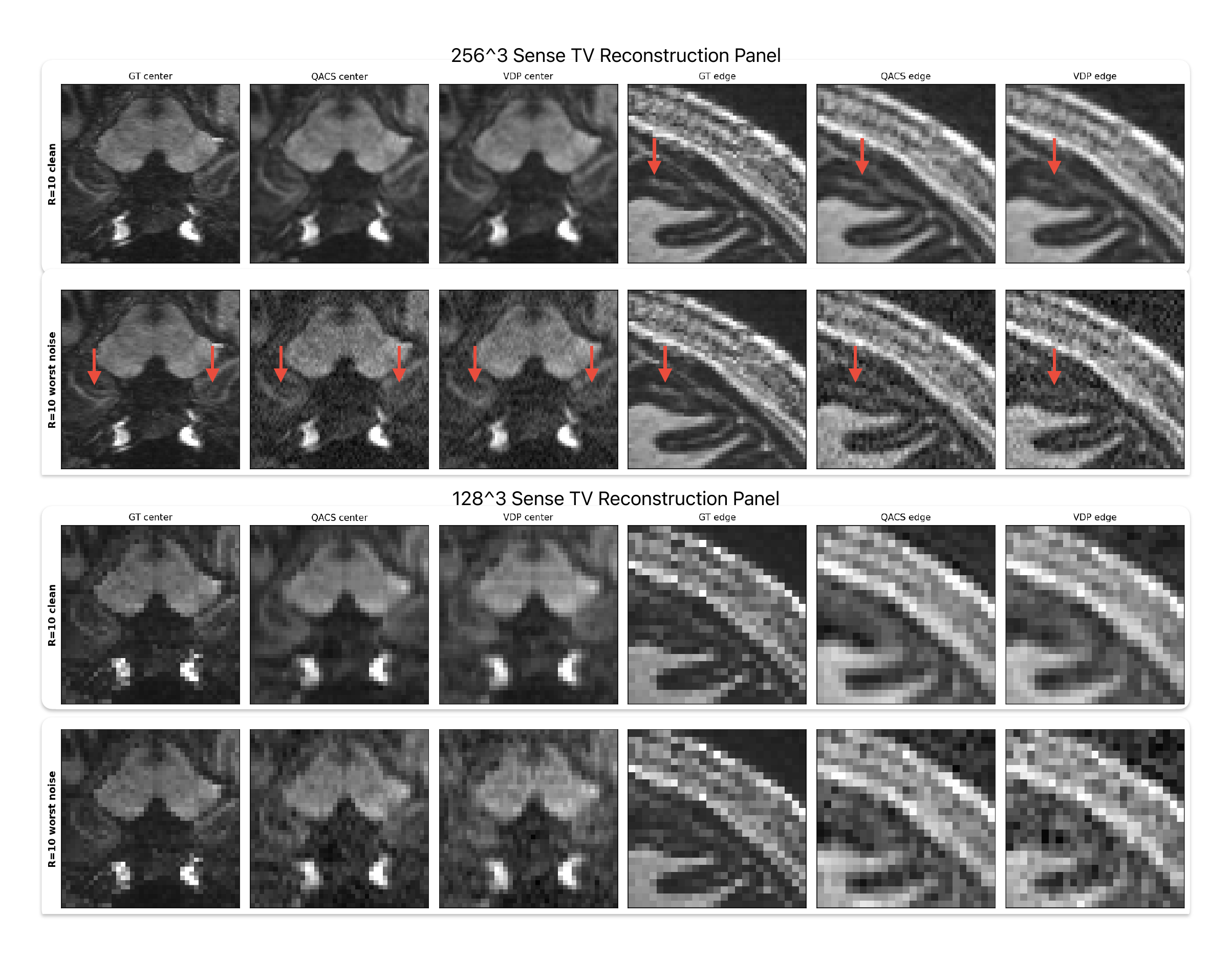}
    \caption{Reconstruction Comparison between QAS and VDP, which are the top two performers amongst all methods. The red arrows indicate that QAS preserves minute phantom details that are degraded when using VDP. (Here, GT=Ground Truth)}
    \label{fig:recon_zoom_combined}
\end{figure}

The qualitative reconstruction comparisons for both phantoms further reinforce the quantitative findings (Fig. \ref{fig:recon_combined}). Across both sampling rates, QAS consistently produces visibly sharper visual appearance with improved delineation of cortical structures and fine anatomical details. In noiseless settings, QAS closely matches the ground truth with minimal blurring and reduced aliasing, whereas competing methods, particularly Cartesian VD and spiral sampling, exhibit noticeable smoothing and loss of high-frequency information. Under noisy conditions (20 dB SNR), these differences become more pronounced. QAS maintains clearer tissue boundaries as seen in the red arrowed regions in the Figure:\ref{fig:recon_zoom_combined} and suppresses noise amplification, while VDP and radial methods show increased graininess and partial structural degradation. Spiral and uniform random reconstructions degrade most severely, with significant loss of contrast and structural fidelity. The reduction in ringing artifacts and preservation of edge definition in QAS reconstructions are consistent with its lower HFEN and higher SSIM, highlighting the advantage of the adaptive sampling strategy in capturing and preserving critical k-space information under both clean and noisy conditions.

Furthermore, the center and edge ROI panels highlight improved local reconstruction quality with QAS for both phantoms (Fig. \ref{fig:recon_zoom_combined}). QAS preserves sharper edges and finer details, closely matching the ground truth in clean conditions and maintaining clearer structures under noise. In contrast, VDP shows increased blurring and noise, especially at $R=10$, confirming QAS's superior detail preservation. Figure \ref{fig:recon_zoom_combined} shows that QAS preserves more details of the original phantom than VDP under moderate noise.

\subsection{D'Wave Experiments and results}
\begin{figure}[H]
    \centering
    \includegraphics[width=0.65\textwidth]{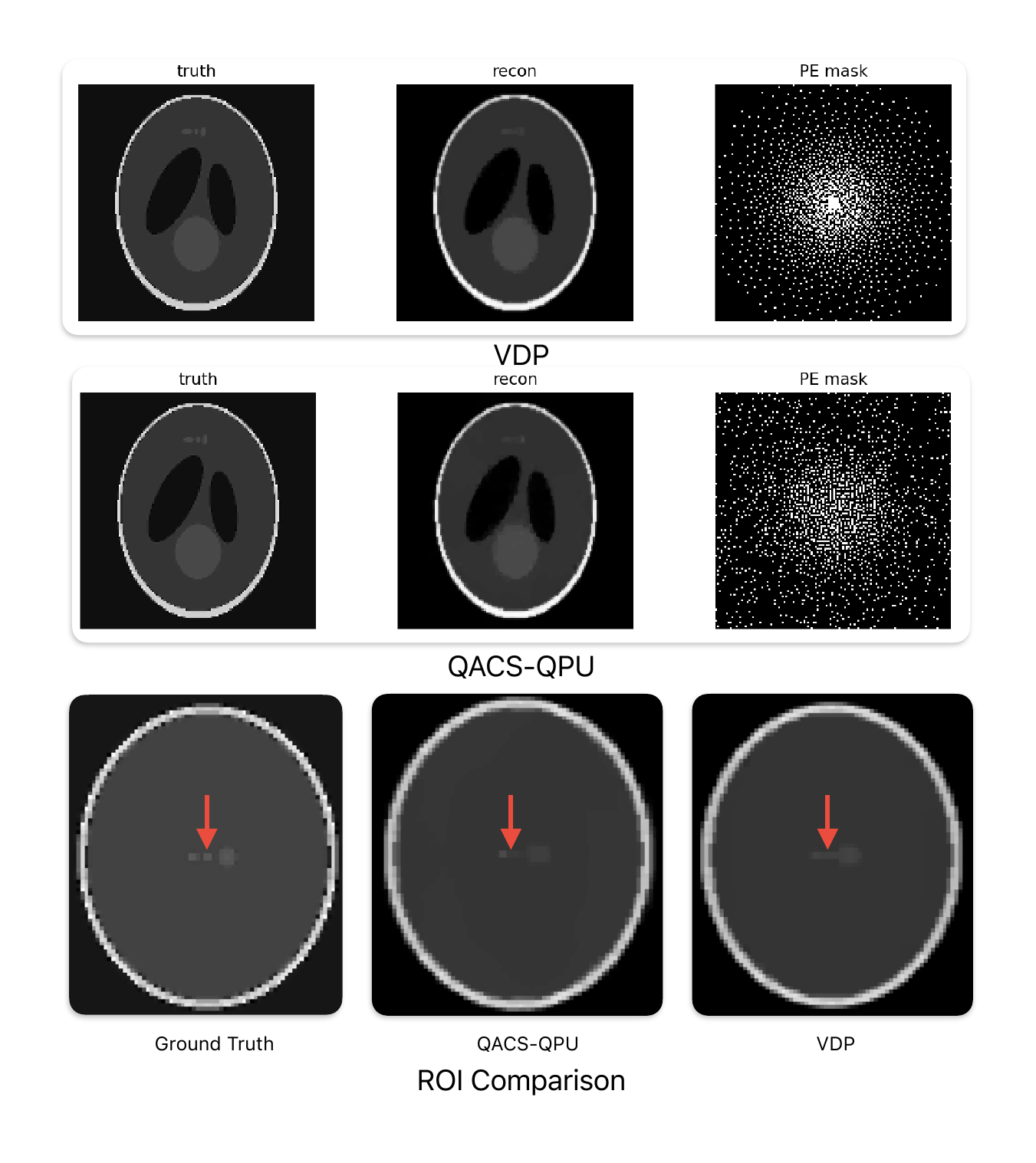}
    \caption{R = 10: Comparison of QAS using quantum annealing and VDP using 8-coil SENSE-TV with 20 dB input noise}
    \label{fig:recon_overview_qpu}
\end{figure}

\begin{table}[H]
\centering
\renewcommand{\arraystretch}{.75}
\begin{tabular}{|l|c|c|c|c|}
\hline
\textbf{Mask at No Input Noise} & \textbf{PSNR} & \textbf{SSIM} & \textbf{NRMSE} & \textbf{HFEN} \\
\hline
Real D-Wave Run Mask & 29.653 & 0.9679 & 0.0533 & 0.1401 \\
\hline
Variable-Density Poisson Mask & 29.845 & 0.9808 & 0.2259 & 0.1291 \\
\hline
\textbf{Mask at 20 dB Input Noise} & \textbf{PSNR} & \textbf{SSIM} & \textbf{NRMSE} & \textbf{HFEN} \\
\hline
Real D-Wave Run Mask & 29.852 & 0.9711 & 0.2258 & 0.1334 \\
\hline
Variable-Density Poisson Mask & 29.928 & 0.9809 & 0.2238 & 0.1255 \\
\hline
\end{tabular}
\caption{Performance metrics comparison at 20 dB input noise and the noiseless setting for $R=10$.}
\label{tab:dwave_mask_performance_20db}
\end{table}
Reconstruction results for the QPU-generated sampling mask are shown in Fig. \ref{fig:recon_overview_qpu}. The resulting sampling mask exhibits a structured yet incoherent distribution, with a denser concentration near the k-space center and dispersed peripheral sampling, consistent with the designed attraction–repulsion formulation. Quantitatively, the SENSE-TV reconstruction achieved the best performance with a PSNR of 29.852 dB. From the experiments done using the real dwave quantum machine we obtained results for the $128^3$ matrix size simulation under noise of 20dB to see the reconstruction under the worst case scenario while the noiseless run is also performed. 
From the results we got for the noisy 20dB run we have the following results in Table:\ref{tab:dwave_mask_performance_20db}.

From the metrics and the obtained slices in Figure:\ref{fig:recon_overview_qpu}, we can see that the image reconstructions are very comparable between QAS and VDP. This is due to the fact that we are using the pooling\footnote{Please refer to details in \ref{pool}} variant of the algorithm, where the pool size is smaller than the total sample size, resulting in a suboptimal mask. This decision was made because the computation for the D'Wave exceeded the desired usage and reduced overhead for the more realistic $128^3$ run.
We can see even with the issues, we can spot improvements in the reconstruction in the very sensitive regions in Figure \ref{fig:recon_overview_qpu} where in the VDP completely joins the 3 small structure as pointed out in the aforementioned figure.

\begin{figure}[H]
    \centering
    \includegraphics[width=0.75\textwidth]{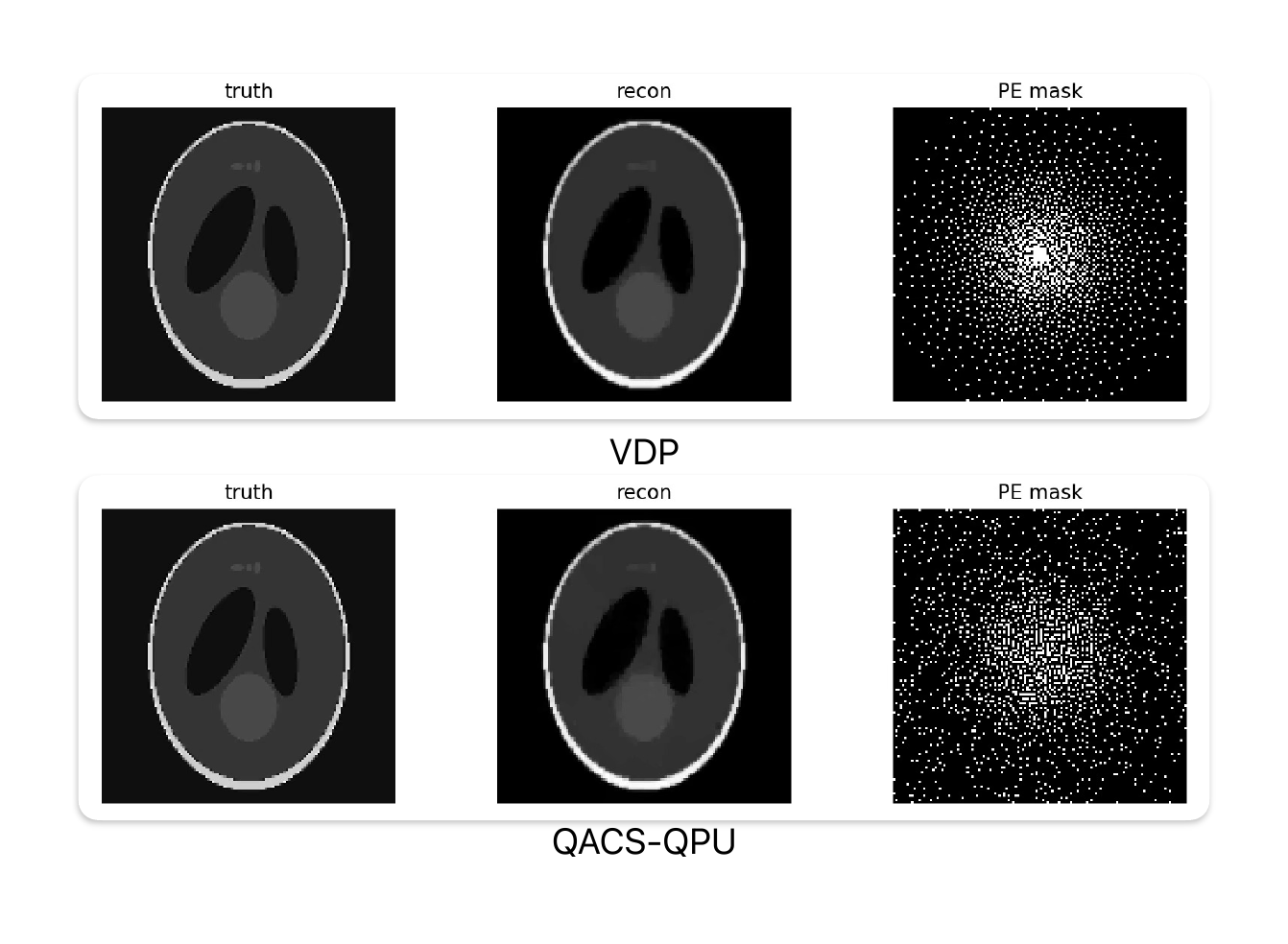}
    \caption{R = 10 (10\% sampling): Comparison of QAS using quantum annealing and VDP using 8-coil SENSE-TV under noiseless setting}
    \label{fig:recon_overview_qpu_noiseless}
\end{figure}

In a noiseless situation, QAS again produced comparable reconstructions to VDP owing to its limited pooling method. Figure \ref{fig:recon_overview_qpu_noiseless} shows the reconstructions obtained by QAS and VDP.

\section{Discussion}

This study demonstrates that adaptive QUBO-based sampling can improve accelerated MRI reconstruction relative to static Cartesian undersampling strategies. Across both $128^3$ and $256^3$ simulations, QAS achieved the best overall reconstruction quality at sampling rates of 20\% and 10\%, with improvements in PSNR, SSIM, NMSE, and HFEN. The benefit was maintained under simulated measurement noise and was generally more apparent when the sampling budget was limited. The reduced-pool D-Wave experiment further demonstrated that the proposed optimization can be executed using current quantum--classical infrastructure, although it did not establish a quantum computational advantage.

\subsection{Why adaptive sampling improves reconstruction}

The performance of QAS can be explained by the complementary roles of the static, adaptive, and pairwise components of its objective function. Conventional variable-density masks encode a population-level assumption that central k-space should be sampled more densely, but the mask remains fixed and cannot respond to the signal measured during an individual acquisition. In contrast, QAS uses previously acquired data to update the attraction field and preferentially sample regions associated with greater measured signal energy. The sampling distribution is therefore progressively conditioned on the object being imaged rather than determined entirely before the scan.

The adaptive field alone could over-concentrate samples around a small number of high-energy locations. QAS avoids this behavior through its pairwise repulsion and exploration terms. The repulsive interaction discourages the simultaneous selection of nearby phase-encoding locations, thereby promoting spatially dispersed and relatively incoherent coverage. The exploration bonus additionally favors the frontier of the acquired region and reduces the likelihood that the optimization repeatedly exploits only central k-space. Log compression, spatial diffusion, and hyperbolic-tangent suppression limit the influence of exceptionally bright measurements and allow neighboring, not-yet-observed locations to inherit information from acquired samples. Consequently, the optimized mask balances three competing requirements: preservation of low-spatial-frequency signal, adaptation to object-specific information, and sufficiently dispersed sampling for compressed-sensing reconstruction.

This balance is consistent with the quantitative results. The reductions in NMSE indicate improved global reconstruction accuracy, whereas the lower HFEN and higher SSIM suggest better preservation of edges and fine spatial structure. The advantage over uniform random sampling also shows that incoherence alone is insufficient. Similarly, the improvement over variable-density Poisson-disc sampling suggests that combining spatial dispersion with measurement-dependent feedback can be more effective than applying a static variable-density distribution.

\subsection{Advantages of the QUBO formulation}

Formulating batch selection as a quadratic unconstrained binary optimization problem provides several practical advantages. First, the binary variables provide a natural representation of whether each candidate phase-encoding line is selected. Second, the linear coefficients can encode the center preference, measured signal energy, and exploration bonus, while the quadratic coefficients directly represent interactions between candidate lines. The fixed sampling budget is incorporated through the quadratic penalty
\[
P\left(\sum_i s_i-M\right)^2,
\]
which allows the attraction, dispersion, and cardinality requirements to be optimized jointly instead of enforcing them through separate heuristic steps.

The QUBO representation is also solver-independent. The same objective can be addressed using parallel tempering, simulated annealing, specialized classical QUBO solvers, quantum annealers, or quantum--classical hybrid algorithms without changing the underlying sampling model. This property allowed the main experiments to be conducted using parallel tempering while retaining compatibility with the D-Wave implementation. The principal methodological contribution should therefore be interpreted as the adaptive optimization framework and its QUBO encoding, rather than as a claim that the observed reconstruction improvements arise uniquely from quantum computation.

Another advantage is that a batch of phase-encoding lines is selected jointly. A greedy method that chooses only the highest-attraction candidate at each step may repeatedly select spatially redundant locations and can become trapped in a locally favorable sampling pattern. Joint optimization accounts for how every selected line interacts with the other lines in the batch. This makes it possible to select a set whose combined information and spatial coverage are favorable even when some individual members would not be chosen by a purely greedy rule.
Although signal intensity was used as the guiding heuristic in this study, it can be replaced in a drop-in manner by a more informative information-based criterion to further improve the robustness of the algorithm.
\subsection{Role of quantum annealing and current hardware limitations}

Quantum annealing may become useful for this problem because adaptive sampling produces a sequence of structured combinatorial optimizations that must be solved repeatedly during acquisition. As the spatial resolution, candidate-pool size, number of coils, or batch size increases, the number of binary variables and pairwise interactions also increases. Future annealers with more qubits, higher connectivity, improved coefficient precision, lower noise, and faster input/output pipelines may permit larger candidate pools to be embedded directly. This could reduce the decomposition and pooling overhead currently required and allow the hardware to explore a larger portion of the sampling space within each acquisition update.

Nevertheless, the present results do not demonstrate quantum speedup or quantum advantage with several limitations. The current implementation relies on a hybrid solver, where a significant portion of the optimization is still performed classically, limiting the ability to fully isolate quantum advantage. Additionally, hardware constraints, such as limited qubit connectivity, noise, and embedding overhead, limit the QUBO problem size and require decomposition into smaller batches. Limiting the search space to the pooled \ref{pool} k-space lines also yields a suboptimal sampling mask with the D'Wave Hybrid solver. This likely contributed to the QPU-derived mask providing reconstruction quality comparable to, rather than consistently better than, variable-density Poisson-disc sampling. The experiments were also conducted on simulated phantom data, and further validation on real MRI acquisitions is required. 

Hardware noise, finite coupling precision, chain breaks, queueing, data-transfer latency, and hybrid-solver overhead are additional obstacles to low-latency scanner integration. Moreover, comparisons between quantum and classical approaches must include the complete wall-clock workflow, including QUBO construction, embedding, communication, solver access, and post-processing, rather than reporting annealing time alone. Systematic benchmarking against parallel tempering and modern classical combinatorial optimizers is therefore necessary before conclusions can be drawn regarding computational advantage.

\subsection{Clinical translation}

The current study is a retrospective proof of concept based primarily on simulated volumetric phantom data, analytical eight-coil sensitivity maps, and SENSE-TV reconstruction. Clinical translation will require prospective validation using raw multi-coil measurements from human participants and physical phantoms. Such studies should include different anatomies, contrasts, field strengths, receive arrays, resolutions, and acceleration factors. Evaluation should also extend beyond global image-similarity metrics to clinically relevant endpoints, including lesion conspicuity, quantitative-map accuracy, diagnostic confidence, and robustness to motion, flow, off-resonance, and coil-sensitivity estimation errors.

The proposed Cartesian line-selection framework is favorable for translation because it can, in principle, be incorporated into conventional phase-encoding acquisitions without requiring a completely new non-Cartesian readout. 

\subsection{Toward real-time scanner implementation}

In the $256^3$ experiments, parallel tempering required approximately $214$--$346$~ms per batch, while the complete QAS optimization time was substantially smaller than the offline SENSE-TV reconstruction time. These measurements indicate that mask optimization was not the dominant computational cost in the retrospective pipeline. They do not, however, establish real-time feasibility because a prospective implementation must complete each feedback update within the timing constraints of the pulse sequence.

A practical scanner implementation could begin with a fully sampled central calibration region or a small initial batch. Immediately after each batch is acquired, the scanner would calculate coil-combined line energies, update the diffused adaptive attraction field, construct the QUBO for the remaining candidates, and solve for the next batch while the current acquisition and data transfer are proceeding. The selected phase encodes would then be inserted into the sequence queue. QUBO construction and optimization could be executed asynchronously on a local CPU, GPU, FPGA, or dedicated optimization accelerator so that computation overlaps with acquisition. Warm-starting the solver from the previous solution and updating only coefficients affected by the newest measurements could further reduce latency.

Initial prospective implementations should use a classical solver because it avoids remote communication and queueing delays and can provide deterministic upper bounds on update time. Quantum or hybrid backends could subsequently be substituted through the same QUBO interface when their end-to-end latency and accessible problem size become compatible with scanner timing. If an optimization result is not returned before a predefined deadline, the scanner should continue with a precomputed variable-density fallback batch. This design would preserve acquisition reliability while allowing adaptive decisions whenever they are available.

\section{Conclusion}

The proposed methodology demonstrated that images can be recovered from randomly under-sampled data, provided a nonlinear recovery scheme is used to enforce data consistency unlike the traditional methods of reconstructing images from under-sampled k-space data utilizing Compressed Sensing to exploit the transform sparsity of images within wavelet or finite-difference domains. Comparisons with other sampling algorithms show that this framework has a higher capability of producing masks that have excellent reconstruction even at low sampling rates. This method even shows significantly better results under moderate noise where most algorithms struggle to produce useful images. On the other hand, this algorithm makes a shift towards active sampling with dynamic optimization, utilizing quantum optimization techniques to navigate complex energy landscapes and give the most optimal sampling mask.

\section{Acknowledgments}

We acknowledge Dr. Daniel Lidar at Quantum Computation and Open Quantum Systems, University of Southern California, for his help in D’Wave’s implementation.

\bibliography{references}

\end{document}